\documentclass{article}

\usepackage{microtype}
\usepackage{graphicx}
\usepackage{subfigure}
\usepackage{caption}
\usepackage{subcaption}
\usepackage{booktabs} 

\usepackage{hyperref}

\newcommand{\bfrt}{\bm{r},t}

\newcommand{\ubr}{\underline{\bm{r}}}

\newcommand{\s}{_\mathrm{{\scriptscriptstyle S}}}
\newcommand{\h}{_\mathrm{{\scriptscriptstyle H}}}
\newcommand{\xc}{_\mathrm{{\scriptscriptstyle XC}}}

\newcommand{\ext}{_\mathrm{{\scriptscriptstyle ext}}}

\usepackage[accepted]{icml2024}

\usepackage{amsmath}
\usepackage{amssymb}
\usepackage{mathtools}
\usepackage{amsthm}
\usepackage{bm}
\usepackage[capitalize,noabbrev]{cleveref}

\theoremstyle{plain}

\theoremstyle{definition}

\theoremstyle{remark}

\usepackage[textsize=tiny]{todonotes}

\icmltitlerunning{Accelerating Electron Dynamics Simulations through Machine Learned Time Propagators}

\begin{document}

\twocolumn[
\icmltitle{Accelerating Electron Dynamics Simulations through Machine Learned Time Propagators}




\begin{icmlauthorlist}
\icmlauthor{Karan Shah}{casus,hzdr}
\icmlauthor{Attila Cangi}{casus,hzdr}

\end{icmlauthorlist}

\icmlaffiliation{casus}{Center for Advanced Systems Understanding, G\"orlitz, Germany}
\icmlaffiliation{hzdr}{Helmholtz-Zentrum Dresden-Rossendorf, Dresden, Germany}

\icmlcorrespondingauthor{Attila Cangi}{a.cangi@hzdr.de}

\icmlkeywords{Machine Learning, ICML}

\vskip 0.3in
]



\printAffiliationsAndNotice{}  

\begin{abstract}
Time-dependent density functional theory (TDDFT) is a widely used method to investigate electron dynamics under various external perturbations such as laser fields. In this work, we present a novel approach to accelerate real time TDDFT based electron dynamics simulations using autoregressive neural operators as time-propagators for the electron density. By leveraging physics-informed constraints and high-resolution training data, our model achieves superior accuracy and computational speed compared to traditional numerical solvers. We demonstrate the effectiveness of our model on a class of one-dimensional diatomic molecules. This method has potential in enabling real-time, on-the-fly modeling of laser-irradiated molecules and materials with varying experimental parameters.
\end{abstract}

\section{Introduction}
Time-Dependent Density Functional Theory (TDDFT) \cite{runge_density-functional_1984} is a widely used method to study the evolution of electronic structure under the influence of time-dependent potentials. It is used to calculate various excited state properties such as excitation energies \cite{adamo_calculations_2013}, charge transfer \cite{maitra_charge_2017}, stopping power \cite{yost_examining_2017}, optical absorption spectra \cite{jacquemin_excited-state_2011} and non-linear optical properties \cite{goncharov_non-linear_2014}. Due to its favorable balance between accuracy and computational cost, TDDFT has been applied in various domains including photocatalysis, biochemistry, nanoscale devices and the study of light-matter interactions in general.

For weak perturbations, the linear response (LR) formalism of TDDFT is used to calculate the excitation spectrum of a system. It is calculated using the Casida equation \cite{casida_time-dependent_1995} as an eigenvalue problem that describes the first order response of the density. In contrast, the electron density is directly propagated in time under the real time (RT) formalism. RTTDFT can be used to calculate the nonlinear response of the density under strong perturbations such as ultrafast electron dynamics with strong laser fields.

There are multiple components that must be decided for RTTDDFT calculations. These include preparation of the initial density and orbitals, choice of the exchange-correlation functional, the form of the time-dependent potential and the choice of the time-propagation scheme. The time evolution is a significant fraction of the computation required \cite{castro_propagators_2004, gomez_pueyo_propagators_2018}.

Machine learning (ML) for accelerating scientific simulations is a rapidly growing area of research \cite{carleo_machine_2019}. A variety of unsupervised and data-driven models have been developed to solve differential equations across a wide range of domains \cite{karniadakisPhysicsinformedMachineLearning2021a}. 

Neural operators (NOs) \cite{kovachki_neural_2023}
are a class of models that map function-to-function spaces, as opposed to the finite-dimensional vector space mappings of neural networks. This is especially useful for partial differential equation (PDE) problems where experimental or simulation data is available. Fourier neural operators (FNOs) \cite{liFourierNeuralOperator2020} are a type of NO which represent operator weights in Fourier space. The main advantages of FNOs are that they generalize well across function spaces, and being resolution-invariant, they can be used for inference on higher resolution grids than the training set grids. FNOs have also been used for forward and inverse PDE problems in various domains \cite{azizzadenesheli_neural_2024}.

While many applications of ML have been developed for ground state DFT \cite{snyder_finding_2012, brockherde_bypassing_2017, fiedler_deep_2022}, there have been relatively fewer efforts focused on TDDFT. Some applications of ML for TDDFT include development of exchange-correlation potentials \cite{yang_machine-learning_2023, suzuki_machine_2020} and predicting properties such as spectra and stopping power \cite{ward_accelerating_2023}. In this work, we demonstrate the effectiveness of FNOs in propagating the electron density in time under the TDDFT framework. Instead of propagating orbitals in time as done conventionally in terms of the time-dependent Kohn-Sham equations, we use the FNO propagator to directly evolve the density. This has two advantages: the computational cost does not scale with the number of orbitals and larger propagation time steps can be used, thus using fewer iterations. Section 2 contains a brief description of TDDFT, FNOs and our proposed autoregressive model. Section 3 contains results obtained with this model for the time evolution of one-dimensional diatomic molecules under the influence of an oscillating laser pulse. We show that the model can be generalized across ionic configurations, is faster and more accurate than comparable numerical simulations and can also be used for higher resolution grids. In Section 4, we discuss the physical viability of the predicted densities, and then offer conclusion in Section 5.

We use atomic units a.u. ($\hbar = m_e = e = 1)$ unless specified otherwise.
\section{Methods}


\begin{figure*}
    \centering
    \subfigure[Ground State Density]{\includegraphics[width=0.3\textwidth]{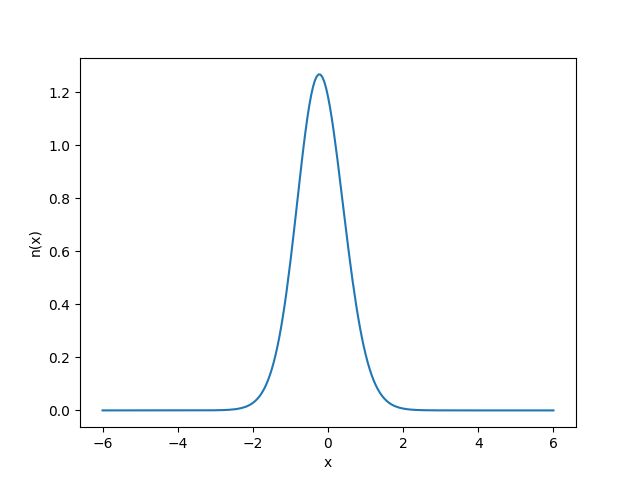} \label{fig:gs_dens}}
    \subfigure[Ground State Potentials]{\includegraphics[width=0.3\textwidth]{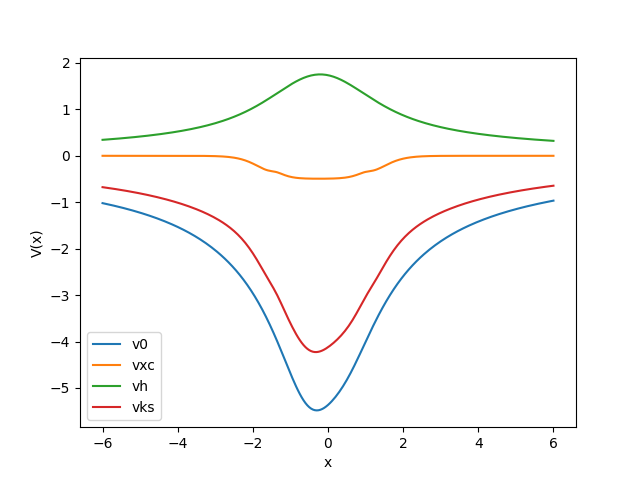} \label{fig:gs_pot}}
    \subfigure[Laser Field]{\includegraphics[width=0.3\textwidth]{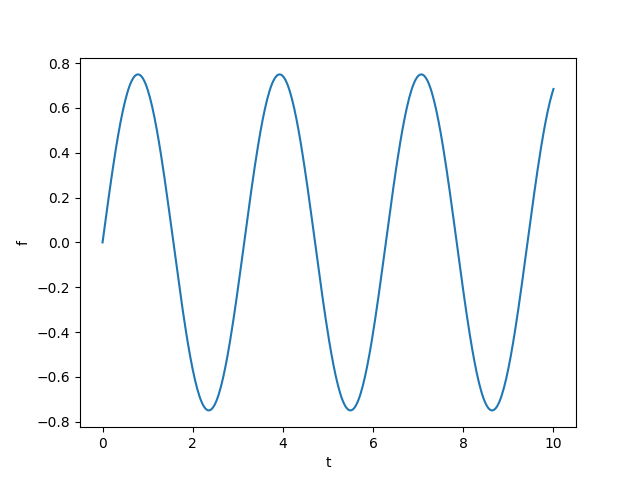} \label{fig:laser}}
    \caption{Example system with $d=1.0, Z_1=2.0, Z_2=4.0$. (a) Ground State Density, (b) Ground State Potentials, (c) Laser Field}
    \label{fig:example_system}
\end{figure*}

\subsection{Time-Dependent Density Functional Theory}
The time-dependent many-body Schr\"odinger equation describes the dynamics of systems composed of multiple interacting particles. A common problem is interacting electrons in atoms and molecules, for which the Schr\"odinger equation is
\begin{equation}
i \frac{\partial}{\partial t} \Psi(\ubr, t)=\hat{H}(t) \Psi(\ubr, t),
\label{eq:tdse}
\end{equation}
where $\ubr$ represents the collective coordinates of the $N$ electrons, $\Psi(\ubr, t)$ is the many-body wavefunction and $\hat{H}$ is the Hamiltonian operator, representing the total energy of they system. For this problem the form of $\hat{H}$ is $\hat{H}(t)=\hat{T}+\hat{W}+\hat{V}(t)$, where $\hat{T}=\sum_{j=1}^N-\nabla_j^2/2$ is the kinetic energy operator, $\hat{W}=\frac{1}{2} \sum_{\substack{j, k \\ j \neq k}}^N 1/\left|\mathbf{r}_j-\mathbf{r}_k\right|$ represents the electron-electron interaction, and $\hat{V}(t)=v\ext(\bfrt)$ is the time-dependent external potential operator which includes the static external potential due to the ions and an external time-dependent potential that drives the system. Note that in writing this Schr\"odinger equation we are assuming the Born-Oppenheimer approximation, i.e. we are treating the dynamics of the ions as occurring on a much longer time scale than the dynamics of the electrons, and can therefore treat the ions as classical point-like particles. The initial state $\Psi(\ubr) = \Psi(\ubr, t_0)$ is obtained by solving the time-independent Schr\"odinger equation
\begin{equation}
    \hat{H}\Psi(\ubr) = E \Psi(\ubr)
    \label{eq:tise}
\end{equation}
which is an eigenvalue problem where $E$ denotes the eigenvalues and corresponds to the total energy of the system.

The many-body wavefunction $\Psi$ contains all the information about the system and can be used to calculate its properties. However for 3 spatial dimensions, we need to solve a system of $3N$ variables and the computational cost scales exponentially with $N$, rendering it intractable for all but the simplest systems. DFT and TDDFT allow us to solve Eq. \ref{eq:tise} and Eq. \ref{eq:tdse} respectively by reformulating the problem in terms of electron density rather than the wavefunction, dramatically reducing the number of variables from $3N$ to $3$.

The Runge-Gross theorem \cite{runge_density-functional_1984} states that for a system with a given ground state many-body wavefunction $\Psi_0=\Psi_0\left(\boldsymbol{r}, t_0\right)$, there exists a unique mapping between the potential and the time-dependent density. The density can be obtained by solving a system of fictitious non-interacting particles governed by the time-dependent Kohn-Sham (TDKS) equations \cite{van_leeuwen_causality_1998}
\begin{equation}
\left[-\frac{1}{2}\nabla^2 + v_s[n](\bm{r},t)\right]\phi_j(\bm{r},t) = i \frac{\partial \phi_j(\bm{r},t)}{\partial t} , j = 1,..., N
\label{eq:KSequation.TD}
\end{equation}
where the electron density 
\begin{align}
n(\bm{r}, t) = \sum_j \, |\phi_j(\bm{r}, t)|^2, j = 1,..., N
\end{align}
is the quantity of interest. The Kohn-Sham potential $v_s[n](\bm{r},t)$ is a functional of the density $n$ which is a function of position $\bm{r}$ and time $t$. The correspondence between density and potential is established through the Kohn-Sham potential
\begin{align}
v\s[n](\bfrt) = v\ext(\bfrt) + v\h[n](\bfrt) + v\xc[n](\bfrt).
\end{align}
where $v\ext(\bfrt)$ is the external potential, $v\h(\bfrt)$ is the Hartree potential  and $v\xc(\bfrt)$ is the exchange-correlation potential. The external potential is composed of the potential of the ions and any time-dependent external perturbation such as a laser field.

While in theory, $v\xc(\bfrt)$ depends on the density at all previous time steps, the adiabatic approximation is often used in practice \cite{ullrich_time-dependent_2012}. We use the adiabatic local density approximation (ALDA).
Under ALDA, \( v\xc(\bfrt) \) is approximated as:
\begin{equation}
 v\xc^{\text{ALDA}}[n, \Psi_0, \Phi_0](\mathbf{r}, t) =  v\xc^{\text{LDA}}[n](\mathbf{r}, t),
\end{equation}
where \( v\xc^{\text{LDA}} \) is the LDA exchange-correlation potential in ground-state DFT. 

These coupled equations are solved through iterative numerical algorithms which represent a significant computational cost in the TDDFT workflow \cite{castro_propagators_2004}.

\subsubsection{Time Propagators}  \label{tddft_tp}

The general form of the time evolution operator $\hat{U}$ for a time domain $T $ is given by:
\begin{equation}
\phi_i(\mathbf{r}, T) = \hat{U}(T, 0) \phi_i(\mathbf{r}, 0).
\end{equation}
In practice, it is applied multiple times for shorter time periods over the domain
\begin{equation}
\hat{U}(T, 0)=\prod_{i=0}^{N-1} \hat{U}\left(t_i+\Delta t_i, t_i\right)
\end{equation}
with $t_0=0, t_{i+1}=t_i+\Delta t_i$, and $t_N=T$. Additionally, the evolution operator is unitary $\hat{U}^{\dagger}(t+\Delta t, t)=\hat{U}^{-1}(t+\Delta t, t)$, which is required for conserving the electron density, and follows time-reversal symmetry $\hat{U}(t+\Delta t, t)=\hat{U}^{-1}(t, t+\Delta t)$. A time propagation algorithm should account for these properties.
Multiple time propagation methods have been developed for this evolution \cite{castro_propagators_2004, gomez_pueyo_propagators_2018}. For this work, we use the Crank-Nicholson method \cite{Crank_Nicolson_1947} to generate the reference data. We then train a FNO to directly propagate the density in time as opposed to propagating single particle orbitals.

\subsubsection{Model Systems}
We consider an external potential 
\begin{align}
    v_{\text{ext}}(\mathbf{r}, t) = v_{\text{ion}}(\mathbf{r}) + v_{\text{las}}(t)\,,
\end{align}
where we simulate a class of one-dimensional diatomic molecules that serves as a model system with two interacting electrons under the static ionic potential:
\begin{align}
    v_{\text{ion}}(\mathbf{r}) = -\frac{Z_1}{\sqrt{(\mathbf{r}-\frac{d}{2})^2+a^2}} - \frac{Z_2}{\sqrt{(\mathbf{r}+\frac{d}{2})^2+a^2}},
\end{align}
where $Z_1$ and $Z_2$ denote the charge of the atomic wells, $d$ denotes the bond length, and $a$ is a softening parameter for numerical stability.

The system is excited with a laser given by a time-dependent external perturbing potential:
\begin{align}
    v_{\text{las}}(t) = A\sin{\omega t}.
\end{align}

With the dipole approximation, we can treat the laser as spatially constant because the size of our molecule is much small than the wavelength of the laser. Given the ground state prepared with $v_{\text{ion}}(\mathbf{r})$, we propagate the density under the influence of the laser $v_{\text{las}}(t)$.

The domain is defined by [-L, L] for time [0, T] with discretization $\Delta x$ and $\Delta t$.
We use fixed boundary conditions
\begin{align}
    \phi_i(-L, t) = \phi_i(L, t) = 0, \quad \forall t \in [0, T].
\end{align}

The ground state density, ground state potential terms and shape of the laser pulse are shown in in Figure~\ref{fig:example_system}. We use the LDA functional due to its simplicity and accuracy for such simple one-dimensional systems. The density evolution is calculated by solving the time-dependent Kohn-Sham equations with the external potential $v_{\text{ext}}(r, t)$ using the Crank-Nicholson scheme. The numerical simulation is performed using the Octopus RT-TDDFT code \cite{tancogne-dejean_octopus_2020}.

\subsection{Fourier neural operators}  \label{FNO_intro}
\begin{figure*}[!ht]
    \centering
    \includegraphics[width=0.6\textwidth]{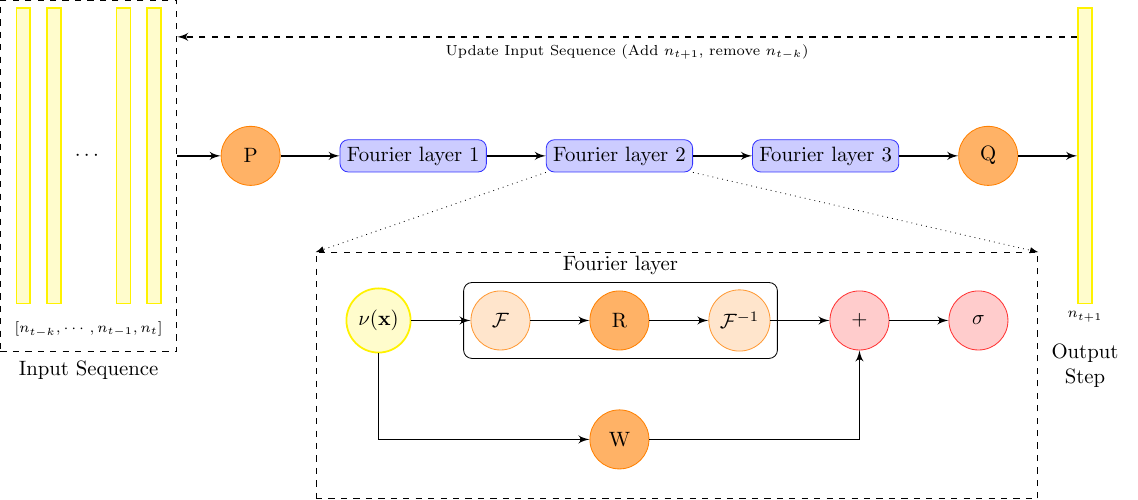}
    \caption{Autoregressive FNO architecture for predicting the density at time $t+1$ based on $k$ previous time steps.}
    \label{fig:FNO_arch}
\end{figure*}

Neural operators (NOs) \cite{luDeepONetLearningNonlinear2019, anandkumarNeuralOperatorGraph2020} extend neural networks by mapping functions to functions instead of finite-dimensional vectors. While a neural network maps $ \mathbb{R}^n $ to $ \mathbb{R}^m $, a neural operator maps function spaces, $ G: \mathcal{A} \rightarrow \mathcal{U} $. Our goal is to approximate the non-linear map $\mathcal{G}^{\dagger}: \mathcal{A} \rightarrow \mathcal{U}$ with a neural operator $\mathcal{G}_\theta$, parameterized by $\theta \in \mathbb{R}^p$.

Training involves observations $ \{ (a_i, u_i) \}_{i=1}^N $ where $ u_i = \mathcal{G}^{\dagger}(a_i) $. The objective is to find parameters $\theta^*$ minimizing the loss:
\begin{equation}
\theta^* = \min _{\theta \in \mathbb{R}^p} \frac{1}{N} \sum_{i=1}^N\left\|u_i-\mathcal{G}_\theta(a_i)\right\|_{\mathcal{U}}^2.
\end{equation}

A neural operator $\mathcal{G}_\theta(a)$ is defined by:
\begin{equation}
\mathcal{G}_\theta(a) = Q(v_L(v_{L-1}(\dots v_1(P(a)))),
\end{equation}
with layers $v_{l+1}$ as:
\begin{equation}
v_{l+1}(x)=\sigma_{(l+1)}\left(W_l v_l(x)+\left(\mathcal{K}_l(a ; \lambda) v_l\right)(x)\right),
\end{equation}
where $\mathcal{K}_l$ is a non-local kernel integral operator:
\begin{equation}
\left(\mathcal{K}_l(a ; \lambda) v_l\right)(x)=\int_{\Omega_l} \kappa_l(x, y,a(x), a(y); \lambda) v_l(y) dy.
\end{equation}

The kernel function $\kappa_l(x,y,a(x), a(y))$ depends on $x, y, a(x)$, and $a(y)$, and is parameterized by $\lambda$. $W_l$ is a learnable weights matrix which corresponds to a linear transformation and $\sigma_{l+1}$ is a component-wise non-linear activation function. $P(a)$ preprocesses $a$ into higher dimensional space $v_0$, and $Q(v_L)$ post-processes $v_L$ back into $\mathcal{U}$.

FNOs \cite{liFourierNeuralOperator2020} use the Fourier transform for efficient computation. With $\kappa(x-y)$ as a convolution operator, we use the convolution theorem:
\begin{equation}
\left(\mathcal{K}_l(a ; \lambda) v_l\right)(x)=\mathcal{F}^{-1}\left(\mathcal{F}\left(\kappa_l\right) \cdot \mathcal{F}\left(v_l\right)\right)(x),
\end{equation}
where $\mathcal{F}$ and $\mathcal{F}^{-1}$ are the Fourier transform and its inverse, respectively. Parameterization in Fourier space is determined by the number of modes $k_{max}$ and width of the convolutional layers, allowing efficient kernel computation and effective capture of global patterns using Fast Fourier Transform.

\subsection{Modeling the time-dependent electron density with Fourier neural operators}  \label{FNO_time}
An autoregressive model is used to predict future values based on past observations. The model takes a slice of $ T_{\text{in}} $ time steps and predicts the next time step. The input sequence is then updated to include the new predicted value while discarding the oldest time step.

Let $ \mathbf{n}_{t} $ be the density grid at time step $ t $. The input to the model at time step $ t $ is a sequence of $ T_{\text{in}} $ previous observations, denoted as:
\begin{equation}
\mathbf{N}_t = \left[ \mathbf{n}_{t-T_{\text{in}}+1}, \mathbf{n}_{t-T_{\text{in}}+2}, \ldots, \mathbf{n}_{t} \right],
\end{equation}

with $\mathbf{N}_0 $ consisting of the initial $T_{in}$ time steps beginning with the ground state density $\mathbf{n}_{0}$. 

The model predicts the next time step $ \hat{\mathbf{n}}_{t+1} $ as:
\begin{equation}
\hat{\mathbf{n}}_{t+1} = \mathcal{G}_\theta(\mathbf{N}_t)
\end{equation}
where $ \mathcal{G}_\theta(\cdot) $ represents the autoregressive FNO model parameterized by $ \theta $.

After predicting $ \hat{\mathbf{n}}_{t+1} $, the input sequence is updated for the next prediction. The new input sequence $ \mathbf{N}_{t+1} $ is formed by appending $ \hat{\mathbf{n}}_{t+1} $ and removing the oldest time step $ \mathbf{n}_{t-T_{\text{in}}+1} $:
\begin{equation}
\mathbf{N}_{t+1} = \left[ \mathbf{n}_{t-T_{\text{in}}+2}, \mathbf{n}_{t-T_{\text{in}}+3}, \ldots, \mathbf{n}_{t}, \hat{\mathbf{n}}_{t+1} \right]
\end{equation}

This architecture is shown in Figure~\ref{fig:FNO_arch}.
This process is repeated iteratively to generate further predictions. 

The loss function for this model is the mean squared error (MSE) between the predicted and true densities, averaged over all time steps and all systems. Let $\mathcal{D}$ denote the set of all systems, and let $\mathcal{T}$ denote the set of all predicted time steps for a given system. The loss function $\mathcal{L}(\theta)$ is defined as:
\begin{equation}
\mathcal{L}(\theta) = \frac{1}{|\mathcal{D}|} \sum_{d \in \mathcal{D}} \frac{1}{|\mathcal{T}|} \sum_{t \in \mathcal{T}} \left\| \mathbf{n}_{t}^{(d)} - \hat{\mathbf{n}}_{t}^{(d)} \right\|_2^2,
\end{equation}
where $\mathbf{n}_{t}^{(d)}$ is the true density and $\hat{\mathbf{n}}_{t}^{(d)}$ is the predicted density at time step $t$ for system $d$.

To ensure that the density is conserved through time, we calculate the norm of density

\begin{equation}
    N = \int_{-\infty}^{\infty} n(x, t) dx \approx  \sum_{i} \hat{n}_{t,i}^{(d)} \Delta x 
\end{equation}
\begin{figure*}
    \centering
    \subfigure[Reference Density]{\includegraphics[width=0.24\textwidth]{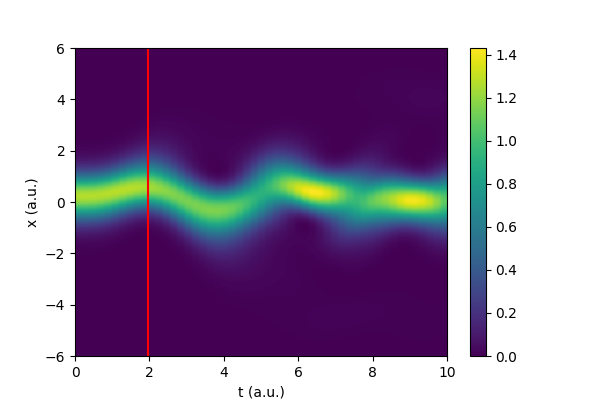} \label{fig:full_snapshot}}
    \subfigure[Predicted Density]{\includegraphics[width=0.24\textwidth]{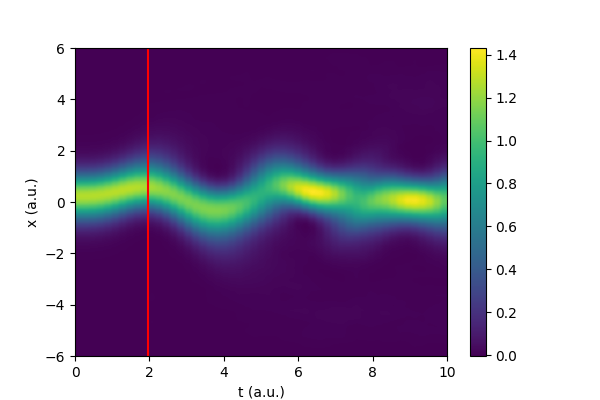} \label{fig:pred_snapshot}}
    \subfigure[AE in Density]{\includegraphics[width=0.24\textwidth]{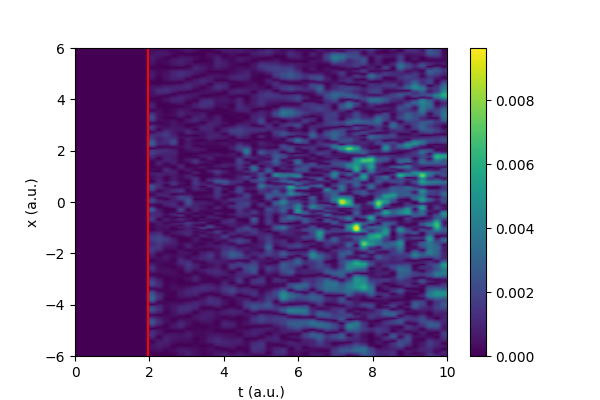} \label{fig:abs_err_snapshot}}
    \subfigure[$log_{10} $(AE) in Density]{\includegraphics[width=0.24\textwidth]{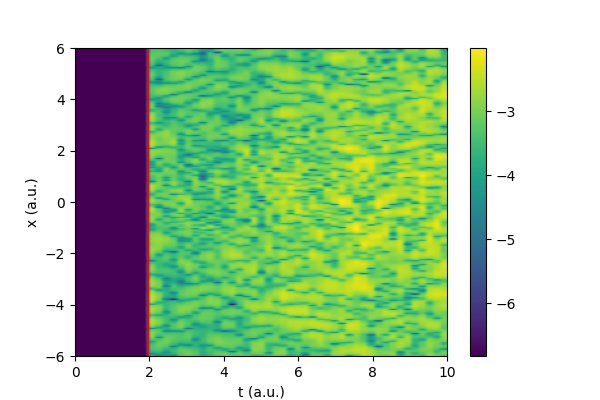} \label{fig:abs_err_log_snapshot}}
    \caption{Density evolution of the representative system. The red line denotes the initial $T_{in}=10$ input time slices. The errors are calculated on the predicted density slices after the red line. }
    \label{fig:snapshots}
\end{figure*}
and add a loss term 
\begin{equation}
    \mathcal{L}_{norm}(\theta) = \lambda \left(  \sum_{i} \hat{n}_{t,i}^{(d)} \Delta x - 2 \right)^2,
\end{equation} where $i$ denotes index over grid points $x_i$ and $\lambda$ is weight term. N = 2 for the reference system.

To evaluate model performance, we use the mean absolute error (MAE) and mean squared error (MSE) between the reference and predicted densities.

\section{Results}

As baseline, we generate a set of 729 systems with varying $Z_1, Z_2, d$ parameters with fixed softening parameter $a=1.0$. The domain parameters are $L = 6.0$ a.u. and $T = 10 $ a.u. (0.241 fs). The discretization used is $\Delta x = 0.05$ a.u. and $\Delta t=0.025$ a.u. (0.6 as). The laser parameters are $A = 0.75$ a.u. (corresponding to peak intensity $I=1.97 \times 10^{16}\text{ W cm}^{-2}$) and $\omega = 2.0$ Ha (corresponding to $\lambda = 22.78$ nm). Using this reference set, we generate datasets with discretization $\Delta t=0.2$ a.u. (4.83 as), resulting in $T_n =51$ time slices, each on a grid of $X_n = 241$ spatial points. The training dataset consists of 600 systems, the validation dataset consists of 10 systems and performance is evaluated on a test dataset of 100 systems. The networks consist of 3 Fourier layers, trained using the Adam optimizer with learning rate = 0.001, halving every 200 epochs. 

\subsection{Effect of varying input time slice}
We first show that by incorporating  more information in the input time slices, the performance of the network increases. In this case, the network consists of layers with width 64 and 16 modes. Prediction metrics are evaluated on $T_n-T_{in}$ time slices. As the number of input time slices $T_{in}$ increases, the errors decrease at the expense of the initial data required. We show the error metrics for four models with $T_{in} = {5, 10, 15, 20}$ in Table~\ref{table:performance}. As a balance between the accuracy and data requirements, a larger model is created with $T_{in}=10$, consisting of layers with 16 modes and width 128, trained for 2000 epochs. This model serves as the baseline for further comparisons. The reference and predicted densities along with errors for a representative system are shown in Figure~\ref{fig:snapshots}.

\subsection{Comparison with coarse solver}
We compare the accuracy and speed of the baseline model with a coarse solver. The baseline model is trained on a grid with $\Delta t=0.2$ a.u., using data generated with a finer grid $\Delta t=0.025$ a.u. Numerical results are calculated on a comparable coarse grid $\Delta t=0.2$ a.u. in Octopus. Both these results are compared with reference fine grid numerical calculations. As show in Table ~\ref{table:performance}, the baseline model is roughly $2\times$ faster than the numerical solver with error reductions of about $4\times$ MAE and $10\times$ MSE.

\subsection{Time offset}
We show that the model generalizes well when used to predict the density evolution on a time grid offset by $\frac{\Delta T}{2} = 0.1$ a.u. The domain in this case becomes [$0.1, T-0.1$]. As show in Table ~\ref{table:performance}, the MAE=5.779$\times10^{-3}$ and prediction time per step $t=1.75$ ms is comparable to the performance of the baseline model, which was evaluated on the same time grid as the training time grid.

\subsection{Spatial super-resolution}
One key advantage of FNOs is discretization invariance. We show the performance of the model trained on a dataset with $\Delta x = 0.05$ a.u. ($X_n=241$) on a higher resolution grid with $\Delta x = 0.025$ a.u. ($X_n = 481$). We get accurate predictions with a reasonable increase in errors but with the same inference speed, as shown in Table  \ref{table:performance}. This is especially useful because the time taken per step increases to 9 ms for the numerical solver on this grid while the model inference time stays constant at 1.75 ms.
\begin{table*}[t]\centering

\begin{tabular}{lccc}
\toprule
Model & MAE ($\times 10^{-3}$) & MSE ($\times 10^{-4}$) & Time (ms) \\
\midrule
FNO $T_{in}$ = 05 & 8.595 & 4.18 & 1.58  \\
FNO $T_{in}$ = 10 & 7.087 & 4.10 & 1.62 \\
FNO $T_{in}$ = 15 & 6.101 & 3.16 &  1.60 \\
FNO $T_{in}$ = 20 & 5.652 & 3.32 & 1.60 \\
\midrule
Baseline & 5.118 & 2.56 & 1.75 \\
Octopus (Coarse) & 19.616 & 24.69  & 4.00 \\
\midrule
Time Offset & 5.779 & 2.78 &  1.75\\
\midrule
Super-resolution & 7.335 & 6.56 & 1.75 \\
\bottomrule
\end{tabular}
\caption{Performance metrics of various models. The time column indicates the calculation/inference time to obtain the next density time slice.}
\label{table:performance}

\end{table*}

\section{Discussion}
In this section, we summarize the numerical results and discuss the performance of the FNO time propagator for calculating physically viable densities. For the baseline case, the MAE in density is 5.118 $\times10^{-3}$ with a prediction time of 1.75 ms per time step. This is in excellent agreement to the reference data with a fraction of the computational cost. This model can be used for offset time grids and for higher-resolution space grids as well without additional training. To compare the nature of calculated densities, we evaluate the model on the properties of time evolution operators and use the density to calculate an observable.

\subsection{Properties of time evolution operators}
As discussed in \ref{tddft_tp}, time propagators for the Kohn-Sham equations follow certain properties.
\begin{itemize}
    
    \item Density conservation: 
        The norm of the density $N = \int_{-\infty}^{\infty} n(x, t) dx$ must be conserved in the model system. For our system with two electrons, $N=2$. For the baseline system, the FNO predicted densities have a mean norm of $1.99992 \pm 0.00039$, averaged over all predicted time steps over all systems. The deviation of the predicted norm increases with time, as shown in Figure~\ref{fig:norm_evolution}.

    \item Time reversal symmetry: 
    A stable time-propagator must follow time-reversal symmetry. As a qualitative check, densities are predicted using the baseline model in reverse time order. However, the MAE jumps to 0.0751, with even the MAE on the training set jumping to 0.0741.

The FNO model follows density conservation, which is also constrained in the loss function during training. However, time reversal symmetry is broken as shown by the increase in MAE for reverse time order predictions.

\subsection{Dipole Moment Calculation}
      \begin{equation}
    \mu(t) = \int_{-\infty}^{\infty} x \, n(x,t) \, dx \approx \sum_{i} x_i \, n_{t,i} \, \Delta x
\end{equation}
To demonstrate that the model predicted densities can be used for calculating observables, we also compare the dipole moments calculated using predicted and reference densities. For the baseline, the dipole moment MSE is 0.02976 with a mean average percentage error of 8.34 \%. The error in dipole moment accumulates across time, as shown in Figure~\ref{fig:dm_evolution}.
\begin{figure}[H]
\centering
\includegraphics[width=0.45\textwidth]{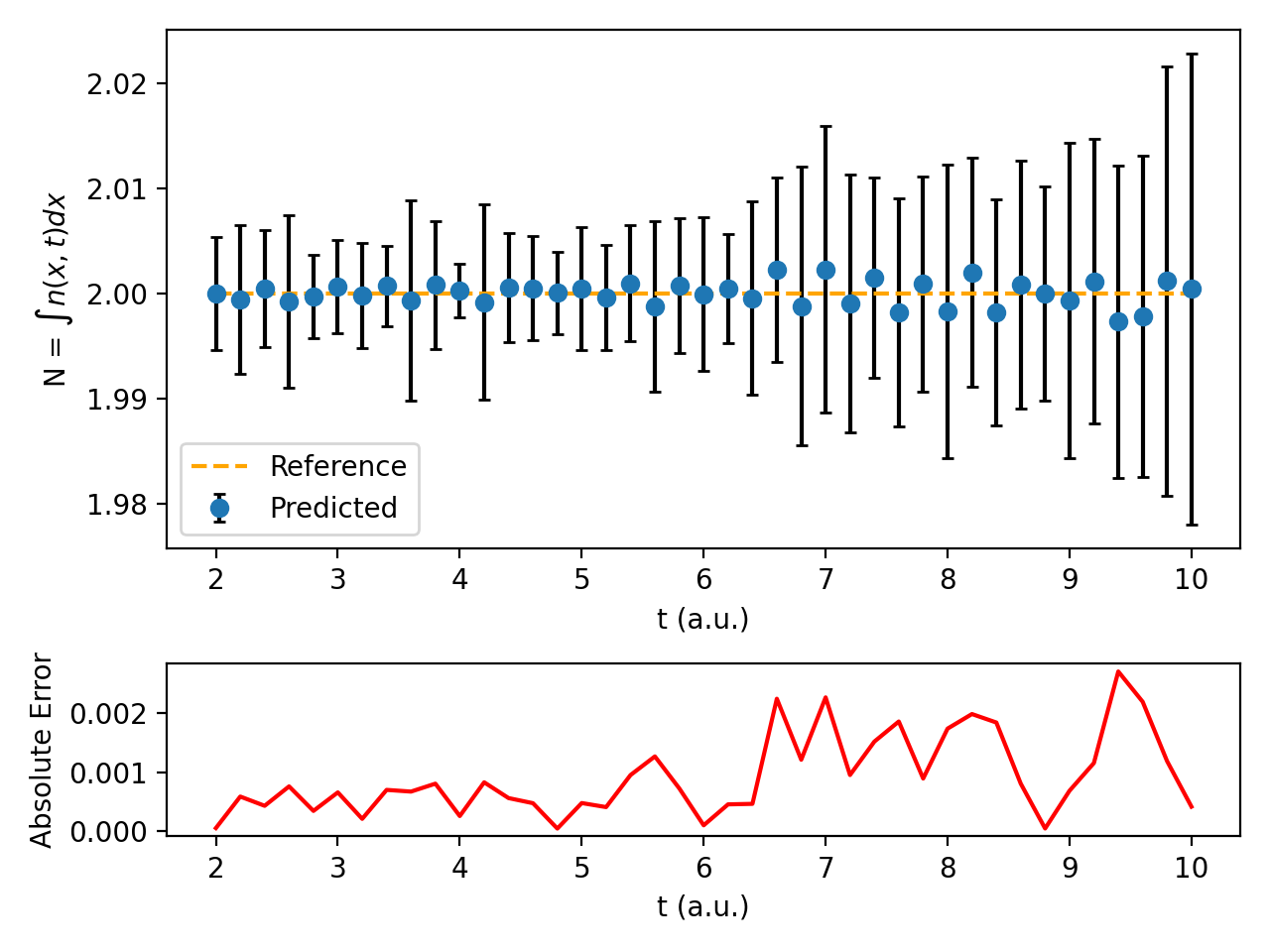}
\caption{Evolution of predicted density norm over time. The blue points represent the norms of predicted density, averaged across all systems, at varying time steps with the error bars representing the stand deviation. The orange line represents the true norm. The subplot shows the absolute error between reference and predicted norm in time.}
\label{fig:norm_evolution}
\end{figure}

\begin{figure}[H]
\centering
\includegraphics[width=0.45\textwidth]{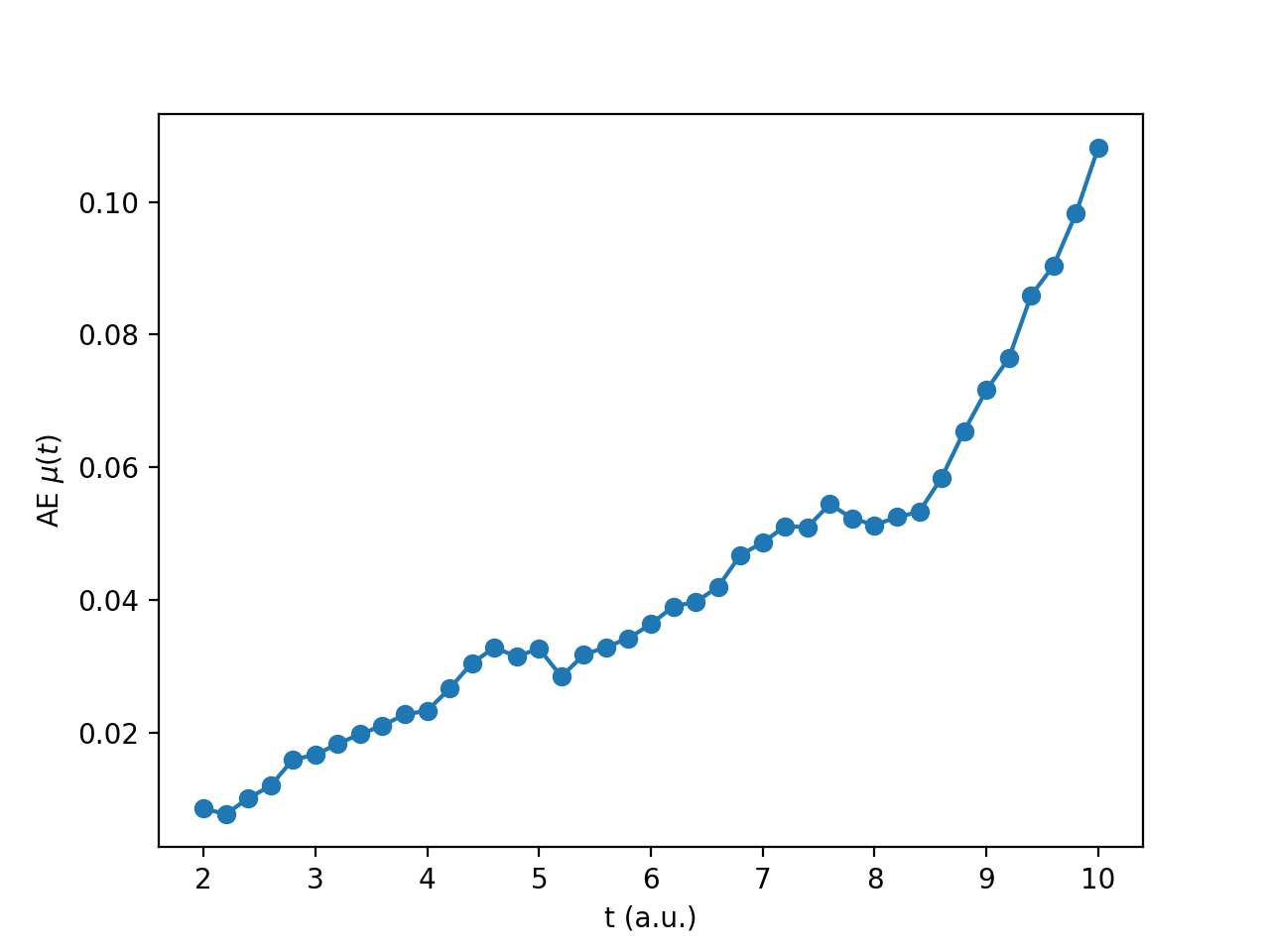}
\caption{Absolute error between the reference and predicted dipole moment, averaged across all systems, at different time steps. }
\label{fig:dm_evolution}
\end{figure}

\end{itemize}
The performance of the FNO time propagator in calculating these physically relevant quantities shows that while density predictions are accurate, the model can be improved further by incorporating more physics-based constraints such as time reversal symmetry and stronger density conservation. Since the typical time scale for electron dynamics is in the range of a few hundred femtoseconds at most, achieving small error accumulation on this time scale seems feasible based on our results.

\subsection{Conclusion}
We show that machine learned time propagators have the potential to accelerate TDDFT calculations. A promising direction would be the development of ML time propagators that can predict the time evolution given just the ground state density and the shape of a laser pulse. Such extensions would include having the laser as a separate input for better generalization across laser parameters and to encode time reversal symmetry in the model structure. Extending this work to physics-informed three-dimensional models would enable on-the-fly modeling of the electronic response properties of laser-excited molecules and materials in various scattering experiments that are conducted at photon sources around the globe. This would enable fast simulations that generalize well over the input parameters of the experimental setup. Rapid modeling would also enable the design of laser pulses to precisely control quantum dynamics under quantum optimal control theory \cite{werschnik_quantum_2007}.

\section*{Acknowledgments}
This work was supported by the Center for Advanced Systems Understanding (CASUS), which is financed by Germany’s Federal Ministry of Education and Research (BMBF) and by the Saxon state government out of the State budget approved by the Saxon State Parliament. We acknowledge funding from the Helmholtz Association’s Initiative and Networking Fund through Helmholtz AI. 

\bibliography{tdks_fno}
\bibliographystyle{icml2024}

\end{document}